\documentclass[%
aip,
jmp,
 amsmath,amssymb,
reprint,%
]{revtex4-2}

\usepackage{graphicx}
\usepackage{dcolumn}
\usepackage{bm}

\usepackage[utf8]{inputenc}
\usepackage[T1]{fontenc}
\usepackage{mathptmx}
\usepackage{color}
\usepackage{etoolbox}
\usepackage{braket}
\graphicspath{{GraphenDiss}}
\usepackage{amsmath}
\usepackage{dsfont}
\usepackage{mathtools}
\usepackage{hyperref}
\usepackage{comment}
\usepackage{amsmath}
\usepackage{amssymb}
\usepackage[paper=a4paper,left=25mm,right=25mm,top=25mm,bottom=25mm]{geometry}
\usepackage{mathtools}
\usepackage{dsfont}
\usepackage{hyperref} 
\usepackage{cleveref}
\crefname{figure}{Fig.}{Fig.}
\crefname{equation}{Eq.}{Eq.}
\usepackage{siunitx}
\usepackage{subcaption}
\usepackage{graphicx} 
\usepackage[sort&compress]{natbib}
\graphicspath{{GraphenDiss}}
\usepackage{float}
\usepackage{orcidlink}

\makeatletter
\def\@email#1#2{%
 \endgroup
 \patchcmd{\titleblock@produce}
  {\frontmatter@RRAPformat}
  {\frontmatter@RRAPformat{\produce@RRAP{*#1\href{mailto:#2}{#2}}}\frontmatter@RRAPformat}
  {}{}
}%
\makeatother
\begin{document}

\preprint{AIP/123-QED}

\title[The Ericson Transition in Time--Reversal Invariant Systems]{The Ericson Transition in Time--Reversal Invariant Systems: Symplectically Invariant Hamiltonians}

\author{Simon Köhnes$^{1}$ and Thomas Guhr$^1$}
\affiliation{
$^1$ Fakult\"at f\"ur Physik, Universit\"at Duisburg--Essen, Duisburg, Germany
             }
\email{simon.koehnes@uni-due.de}

\date{\today}
\begin{abstract}
	We investigate the distributions of off-diagonal scattering matrix elements and cross-sections in the Ericson regime for time--reversal invariant systems with odd spin. Within the universal Heidelberg approach we prove that Gaussian and exponential behavior is found as the leading--order term of an asymptotic expansion of the aforementioned distributions. Furthermore, the subleading--order  term is derived and compared with numerical simulations. 
\end{abstract}
\maketitle
\section{Introduction}
Scattering phenomena lie at the heart of both classical and quantum mechanics. A plethora of examples is found in nuclear
\cite{BRINK196377, VONBRENTANO1964,
	Porter1965,HARNEY199035,Main1992, Feshbach1993,
	Concepts,Weidenmüller, Weidenmüller2,Frisch2014,
	Kawano2015,SmilanskyIrregular}, atomic and molecular physics
\cite{LOMBARDI1993, GREMAUD1993, DUPRET1995, SCHINKE1995,
	Scott1996,
	PhysRevLett.95.194101,PhysRevLett.95.263601,Mayle2013},
quantum graphs \cite{Quantengraphen, QuantumChaos,
	QuantumChaos2,Ławniczak,Celardo2007,Schäfer2003}, microwave
cavities and networks \cite{
	PhysRevLett.65.3072,PhysRevLett.74.62,PhysRevLett.94.144101,
	PhysRevE.74.036213, PhysRevE.81.036205,
	Dietz2010,PhysRevLett.100.254101,Hul2005,Lawniczak2008,Lawniczak2020,Chen2021},
but also classical wave systems \cite{Weaver,PhysRevLett.75.1546,
	GROS2014664}. The theory of electronic transport
\cite{RevModPhys.69.731,PhysRevB.35.1039,RevModPhys.72.895,
	PhysRevLett.69.1584, MIRLIN1994325,
	Folk1996,BringiChandrasekar2001, Yeh2012} is closely
connected. Recently, the field of stochastic quantum scattering  has received considerable attention. Pioneered in a series of seminal papers by Torleif Ericson in the 1960s, the Ericson regime, which is characterised by many channels and strongly overlapping resonances, was first found \cite{Ericson1, Ericson2,
	Ericson3}, see a review in \cite{Weidenmüller2},  and
furthermore in a wide range of systems, \textit{e.g.}, \cite{SmilanskyIrregular}, in electronic
transport \cite{HAW90,SICZB09}, in atoms
\cite{Bluemel96,GD97,PhysRevLett.95.194101,PhysRevLett.95.263601,EM09,EM10,EKM11},
in ultracold atomic-molecular systems~\cite{MRB12}, in
Bose--Einstein condensates \cite{HHCKB12} and quantum graphs
\cite{ZM13}. In the Ericson regime, a universal Gaussian
distribution of the scattering matrix elements emerges, which was proven for the case of time--reversal noninvariant system in \cite{LetterUniversality}. In this work, we extend the analysis to systems with time--reversal invariance and odd spin.
We have the following objectives: First, we explicitly derive the universal Gaussian behavior. Second, we establish the subleading--order corrections to the Gaussian, thereby addressing the transition region where $\Xi \simeq 1$. We also give explicit analytical results for the cross--section distribution and the moments.
Now, we briefly introduce the framework of the Heidelberg approach and general methodology as used in \cite{LetterUniversality}.
The scattering process is described by the unitary scattering matrix $S$
\cite{Mahaux, AGASSI1975,VWZ}. $M$ channels are connected
to an interaction zone. The latter is described by a Hermitean Hamilton
operator, the elements $S_{ab}$ of the scattering matrix being
$S_{ab} = \delta_{ab} - 2\pi i W_a^\dagger G W_b$, where
\begin{align}
	G &=\left( E\mathds{1}_N - H +i\pi\sum\limits_{c=1}^{M} W_c W_c^\dagger\right)^{-1}
\end{align}
is the $N\times N$ matrix resolvent at energy $E$ in a matrix
notation. The limit $N\to \infty$ has to be taken, $H$ is the
Hamilton matrix and the $N$--component vectors
$W_c,\ c=1,...,M$ describe the coupling between bound states
in the interaction zone and the channels. One may assume their
orthogonality.  Apart from kinematic factors, the cross
sections are given by $\sigma_{ab} = |S_{ab}|^2$.
When considering chaotic scattering, $H$ is modeled by a random matrix. This matrix is
drawn from one of the Gaussian Ensembles GOE $(\beta=1)$, GUE
$(\beta=2)$ or GSE $(\beta = 4)$. The cases $\beta= 1,4$ describe
time--reversal invariant systems. Here, we focus on the
time--reversal invariant symplectic $(\beta=4)$ case of real self--dual quaternion matrices $H$ with Gaussian probability distribution
\begin{align}
	P_\beta(H)\propto \exp{\Biggl(-\frac{\beta N}{4\nu^2}\text{tr} H^2\Biggr)}.
\end{align}
The Ericson transition is characterized by the ratio of average resonance
width $\Gamma$ and average mean level spacing $D$, \textit{i.e.} by the
dimensionless parameter
\begin{align}
	\Xi= \frac{\Gamma}{D}= \frac{1}{2\pi} \sum\limits_{c=1}^{M}T_c \ , 
\end{align}
where the second equation is the Weisskopf estimate
\cite{Weisskopf} and $T_c$ are the transmission
coefficients.
The more general Moldauer-Simonius relation is \cite{Simonius,PhysRevC.11.426,DietzWeidenmueller,KottosSchanz,FyodorovJMP1997,FyodorovSommersGamma,PhysRev.123.968,PhysRevC.98.024616,PhysRev.135.B642}
\begin{align}
	\frac{\Gamma}{D}= -\frac{1}{2\pi}\sum\limits_{c=1}^{M}\ln(1-T_c)
\end{align} and was derived for the case of $\beta = 2$ in \cite{FyodorovJMP1997}.
We make use of the supersymmetry method
\cite{Efetov,VWZ,Efetov1996}, using the approach  proposed in
\cite{PhysRevLett.111.030403, Matrixelemente,
	Wirkungsquerschnitte} and that allowed for calculation of the distributions $P_s(x_s)$ of the real and
imaginary parts of $S_{ab}$, denoted by
$x_1=\text{Re}(S_{ab})$, and $x_2=\text{Im}(S_{ab}).$ The distribution of cross section $\sigma_{ab}$ was also derived in terms of
low--dimensional integrals. As in \cite{LetterUniversality}, we use these
results to study the Ericson transition by means of asymptotic
expansions in powers of $1/ \Xi$. 
The upcoming asymptotic analysis is based on the same idea as in \cite{LetterUniversality} and extends the technical methods presented there. Mathematically, in all three Wigner--Dyson classes we asymptotically analyze multidimensional integral representations of characteristic functions exhibiting coordinate singularities. To facilitate explicit asymptotic expansions, it is absolutely crucial to remove these coordinate singularities. Upon doing so, we identify an exponential dependency from the transmission coefficients in each integral, thereby rewriting the characteristic function as a Laplace--type integral. This is then asymptotically evaluated by means of Watson's lemma. After generating the asymptotic expansion, the remaining two--dimensional integrals are evaluated exactly, which is done by mapping the integration region to a rectangle. To elucidate these techniques in detail, we start with the technically less challenging case of $\beta=4$ before we move to the case of $\beta=1$ in the sequel. 
\section{Ericson Transition in the $\beta=4 $ case}
We consider scattering in the Gaussian symplectic case. 
This case corresponds to systems with preserved time--reversal invariance and odd spin. The quantities of interest are the distributions $P_s(x_s)$ of real-- and imaginary parts of $S$--matrix elements  $x_s,\ s=1,\ 2$, where $x_1 = \text{Re}(S_{ambm'})$,  or $x_2=\text{Im}(S_{ambm'})$. The indices $m, m'$ denote the spin orientation associated with the channel $a$ and $b$, respectively. Since the distribution $P_s(x_s)$ is the same for all four possible spin combinations of $m=\uparrow,\downarrow, m'=\uparrow,\downarrow$ \cite{GluthDissertation}, we drop the indices at this point. As in the unitary case, $P_s(x_s)$ is connected to the characteristic function $R_s(k)$ via Fourier transform. 

 The characteristic function for both real and imaginary parts of $S$-matrix elements takes the form \cite{GluthDissertation}
\begin{align}\label{eq:Rsk}
	R_s(k) &=\nonumber 1+\frac{1}{16} \int\limits_{0}^{\infty}d\theta_0 \int\limits_{0}^{\pi/2}d\theta_1 \int\limits_{0}^{\pi}d\theta_2\frac{ \sin{\theta_1}\sin{\theta_2} \sinh{\theta_0}^3 }{(\cos{(\theta_1 - \theta_2) } - \cosh{\theta_0})^2( \cos{(\theta_1+\theta_2)}-\cosh{\theta_0})^2}\\\nonumber&\times\prod\limits_{c=1}^{M} \frac{(g_c^+ + \cos{(\theta_1-\theta_2)})(g_c^+ +\cos{(\theta_1+\theta_2)})}{(g_c^+ + \cosh{\theta_0})^2}\\&\times k^2\left(k(\omega^4+\omega^2t_{ab}+t_{aa}t_{bb})\frac{J_1(\omega k)}{\omega} - 2(2\omega^2+t_{ab})J_0(\omega k)\right),
\end{align}
where
\begin{align}
	t_{aa}&=  \frac{\sin{(\theta_1-\theta_2)}\sin{(\theta_1+\theta_2)}}{(\cos{(\theta_1-\theta_2)}+g_a^+)(\cos{(\theta_1+\theta_2)+g_a^+})},\\ t_{bb}&=\frac{\sin{(\theta_1-\theta_2)}\sin{(\theta_1+\theta_2)}}{(\cos{(\theta_1-\theta_2)}+g_b^+)(\cos{(\theta_1+\theta_2)+g_b^+})},\\ 
	t_{ab}&=\frac{\sin{(\theta_1+\theta_2)}\sin{(\theta_1+\theta_2)}}{(\cos{(\theta_1-\theta_2)}+g_a^+)(\cos{(\theta_1+\theta_2)+g_b^+})}+\frac{\sin{(\theta_1-\theta_2)}\sin{(\theta_1-\theta_2)}}{(\cos{(\theta_1-\theta_2)}+g_a^+)(\cos{(\theta_1+\theta_2)+g_b^+})},\\
	\omega &= \frac{\sinh{\theta_0}}{\sqrt{(g_a^++\cosh{\theta_0})(g_b^++\cosh{\theta_0})}}.
\end{align}

The moments of the corresponding probability distribution are found by examining the coefficients of $k^{2n}, n \in \mathbb{N}_0$ using the series representations of the Bessel functions. 
As seen from \cref{eq:Rsk}, the characteristic function has coordinate singularities at 
\begin{align}
	\cos{(\theta_1+\theta_2)}-\cosh{\theta_0}&=0,\\
	\cos{(\theta_1-\theta_2)}-\cosh{\theta_0}&=0.
\end{align}
These singularities arise from a former parametrization of supermatrices and a singular value decomposition. The integration measure over the saddle-point manifold after integrating out the anticommuting degrees of freedom and commuting degrees of freedom up to the singular values reflects this singular behavior. The singularities, which complicate the explicit evaluation of these integrals, are avoided by choosing new coordinates. We first change to to the basis of singular values $\lambda_j$
\begin{align}
	\theta_0&=\cosh^{-1}{\lambda_0},\\
\theta_1&=\frac{\cos^{-1}{\lambda_1}-\cos^{-1}{\lambda_2}}{2},\\
\theta_2&=\frac{\cos^{-1}{\lambda_1}+\cos^{-1}{\lambda_2}}{2},
\end{align}
The Jacobian of this transformation is 
\begin{align}\left|\frac{\partial(\theta_0,\theta_1,\theta_2)}{\partial(\lambda_0,\lambda_1,\lambda_2)}\right|=\frac{1}{2\sqrt{(\lambda_0^2-1)(1-\lambda_1^2)(1-\lambda_2^2)}}.
\end{align}
While incorporating a factor of $2$ in the integrand, the integration domain in the $\lambda$-basis is then given by
\begin{align}
	1\leq\lambda_0<\infty,\ -1\leq\lambda_1\leq1,\ \lambda_1\leq\lambda_2\leq 1.
\end{align}
Furthermore, we have 
\begin{align}
	\sin{\theta_1}\sin{\theta_2}&=\sin{\left(\frac{\text{arccos}\lambda_1+\text{arccos}\lambda_2}{2}\right)}\sin{\left(\frac{\text{arccos}\lambda_1-\text{arccos}\lambda_2}{2}\right)}= \frac{1}{2}(\lambda_2-\lambda_1),\\\nonumber
	\sinh^3{\text{arccosh}\lambda_0}&=(\lambda_0^2-1)^{3/2}.
\end{align}
Substituting these expressions together with the Jacobian in \cref{eq:Rsk}, we get
\begin{align}
	R_s(k) &\nonumber= 1+\frac{k^2}{16}\int\limits_{1}^{\infty}d\lambda_0\int\limits_{-1}^{1}d\lambda_1\int\limits_{\lambda_1}^{1}d\lambda_2\frac{1}{2}\frac{(\lambda_2-\lambda_1)(\lambda_0^2-1)}{(\lambda_2-\lambda_0)^2(\lambda_1-\lambda_0)^2}
	\frac{1}{\sqrt{(1-\lambda_1^2)(1-\lambda_2^2)}}\prod\limits_{c=1}^{M}\frac{(g_c^++\lambda_1)(g_c^++\lambda_2)}{(g_c^++\lambda_0)^2}\\&\times\left(k(\omega^4+\omega^2t_{ab}+t_{aa}t_{bb})\frac{J_1(\omega k)}{\omega})-2(2\omega^2+t_{ab})J_0(\omega k)\right),\\
	t_{aa}&= \frac{\sqrt{(1-\lambda_1^2)(1-\lambda_2^2)}}{(g_a^++\lambda_2)(g_b^++\lambda_2)},\\
	t_{bb}&=\frac{\sqrt{(1-\lambda_1^2)(1-\lambda_2^2)}}{(g_b^++\lambda_1)(g_b^++\lambda_2)},
	\\
	\omega &= \sqrt{\frac{\lambda_0^2-1}{(g_a^++\lambda_0)(g_b^++\lambda_0)}},\\
	t_{ab}&=\frac{1-\lambda_1^2}{(g_a^++\lambda_1)(g_b^++\lambda_1)}+\frac{1-\lambda_2^2}{(g_a^++\lambda_2)(g_b^++\lambda_2)}.
\end{align}
Now the singularities are at $\lambda_0=\lambda_1=\lambda_2=1$. To make further progress, we change variables again as  
\begin{align}\label{eq:Neue}
	p'&=2\lambda_0 - \lambda_1 - \lambda_2,\\
	s' &= \sqrt{\frac{\lambda_0-\lambda_2}{2\lambda_0-\lambda_1-\lambda_2}},\\
	t'&=\sqrt{\frac{\lambda_0-1}{\lambda_0-\lambda_2}}.
\end{align}
Inversion of \cref{eq:Neue} yields 
\begin{align}
	\lambda_0=1+p's'^2t'^2,\ \lambda_1 = 1-p'(1-s'^2(1+t'^2)),\ \lambda_2 = 1-p's'^2(1-t'^2).
\end{align}
The Jacobian of this transformation evaluates to
\begin{align}
	\left|\frac{\partial(\lambda_0,\lambda_1,\lambda_2)}{\partial(p',s',t')}\right| = 4p'^2s'^3t'.
\end{align}
This transformation is, up to linear transformations, equivalent to the one used in \cite{VerbaarschotCorr} where the correlator of two $S$--matrix elements in the case $\beta=1$ was investigated. The integration region is then given by
\begin{align}
	&\begin{cases}
		0&\leq p' \leq 2, 0\leq s' \leq  \sqrt{\frac{1}{2}}, 0\leq t'\leq 1,\\
		2&\leq p'\leq4, \sqrt{\frac{p'-2}{2p'}}\leq s'\leq \sqrt{\frac{p'-2}{p'}}, \sqrt{\frac{p'(1-s'^2)-2}{p's'^2}}\leq t'\leq1\\
		2&\leq p'\leq4, \sqrt{\frac{p'-2}{p'}}\leq s'\leq \sqrt{\frac{1}{2}}, 0\leq t'\leq1\\
		4&\leq p'<\infty, \sqrt{\frac{p'-2}{2p'}}\leq s'\leq  \sqrt{\frac{1}{2}}, \sqrt{\frac{p'(1-s'^2)-2}{p's'^2}}\leq t'\leq 1.
	\end{cases}
\end{align}
The disconnected part of the manifold, \textit{e.g.}, the asymptotically important one, is given by $0 \leq p' \leq 2, 0\leq s' \leq  \sqrt{\frac{1}{2}}, 0\leq t'\leq 1$. 
In the next step, we write the channel factor 
\begin{align}
		\mathcal{F}_S(\lambda_0,\lambda_1,\lambda_2)&=\prod\limits_{c=1}^{M}\frac{(g_c^++\lambda_1)(g_c^++\lambda_2)}{(g_c^++\lambda_0)^2} 
\end{align}
as exponential and rewrite the occurring logarithms as Taylor series 
\begin{align}
	\mathcal{F}_S&(\lambda_0,\lambda_1,\lambda_2)\\&=\prod\limits_{c=1}^{M}\frac{(g_c^++\lambda_1)(g_c^++\lambda_2)}{(g_c^++\lambda_0)^2} = \exp{\ln{\left(\prod\limits_{c=1}^{M}\frac{(g_c^++\lambda_1)(g_c^++\lambda_2)}{(g_c^++\lambda_0)^2}\right)}}\\&=\exp{\ln{\left(\prod\limits_{c=1}^{M}\frac{(g_c^++1-p'(1-s'^2(1+t'^2)))(g_c^++ 1-p's'^2(1-t'^2))}{(g_c^++1+p's'^2t'^2)^2}\right)}}\\&=\exp{\ln{\left(\prod\limits_{c=1}^{M}\frac{(\frac{2}{T_c}-p'(1-s'^2(1+t'^2)))(\frac{2}{T_c}-p's'^2(1-t'^2))}{(\frac{2}{T_c}+p's'^2t'^2)^2}\right)}}\\&=\exp{\ln{\left(\prod\limits_{c=1}^{M}\frac{(1-\frac{T_c}{2}\ p'(1-s'^2(1+t'^2)))(1-\frac{T_c}{2}\ p's'^2(1-t'^2))}{(1+\frac{T_c}{2}\ p's'^2t'^2)^2}\right)}}
\\&=\exp{\Biggl(\sum\limits_{c=1}^{M}-\frac{p' T_c}{2}\Biggr)}\\&\times	\exp{\Biggl(\sum\limits_{c=1}^{M}\sum\limits_{j=2}^\infty\frac{(-1)^{j+1}}{j}\left(\left(-\frac{T_c}{2}\ p'(1-s'^2(1+t'^2))\right)^j+\left(-\frac{T_c}{2}\ p's'^2(1-t'^2)\right)^j-2\left(\frac{T_c}{2}\ p's'^2t'^2\right)^j\right)\Biggr)}
	\\&\simeq 
	\exp{\left(-\frac{p'(2\pi\Xi)}{2}\right)}(1+\mathcal{O}(p'^2)).
\end{align}
For the leading-- and subleading--order corrections, the integration over $p'$ is done as a Laplace transform. This is done while neglecting exponentially decaying contributions from the $p'=2$ boundary as well as contributions from the connected part of the saddle point manifold. Next, we make use of the recurrence relation of the Bessel functions, 
\begin{align}
	J_{\nu}(x) = \frac{x}{2\nu}\left(J_{\nu+1}(x)+J_{\nu-1}(x)\right).
\end{align} In the $\beta=4$ as well as the $\beta=1$ case, this allows us to rewrite the characteristic function solely in Bessel functions of even order, while in the $\beta=2$ case, there already is only one Bessel function of even order present. Since we want to reconstruct the characteristic function from the asymptotes of the moments, we now derive higher order asymptotes for even-order Bessel functions of the form
\begin{align}
	J_{\nu}\left(k \left(\sum\limits_{j=1}^{\infty}\alpha_{j}(s',t')p'^j\right)^{1/2}\right),
\end{align} where the $\alpha_j(s',t')$ are functions of the remaining integration variables that are determined by a Taylor series of $\omega^2$. Making use of the series of representation of the Bessel functions, we find
\begin{align}
	J_{\nu}\left(k \left(\sum\limits_{j=1}^{\infty}\alpha_{j}(s',t')p'^j\right)^{1/2}\right) = \sum\limits_{n=0}^{\infty}\frac{(-1)^nk^{2n+\nu}}{n!(n+\nu)! 2^{2n+\nu}}\left(\sum\limits_{j=1}^{\infty}\alpha_{j}(s',t')p'^j\right)^{n+\nu/2}.
\end{align} As we are only interested in the leading and subleading order terms in $p'$ of the coefficients of $k^{2n}$ where $p'\ll1$, 
we make the approximation
\begin{align}
	\sum\limits_{j=1}^{\infty}\alpha_{j}(s',t')p'^j\simeq\alpha_{1}(s',t')p'+\alpha_{2}(s',t')p'^2,
\end{align}
and therefore 
\begin{align}
	J_{\nu}\left(k \left(\sum\limits_{j=1}^{\infty}\alpha_{j}(s',t')p'^j\right)^{1/2}\right) \simeq \sum\limits_{n=0}^{\infty}\frac{(-1)^nk^{2n}}{n!(n+\nu)! 2^{2n+\nu}}\left(\alpha_{1}(s',t')p'+\alpha_{2}(s',t')p'^2\right)^{n+\nu/2}.
\end{align}
We proceed by applying a binomial expansion and making further approximations, 
\begin{align}
	\left(\alpha_{1}(s',t')p'+\alpha_{2}(s',t')p'^2\right)^{n+\nu/2} &= \sum\limits_{j=0}^{n}\binom{n+\nu/2}{j}(\alpha_{1}(s',t')p')^{n+\nu/2-j}(\alpha_{2}(s',t')p'^2)^{j}\\&\simeq(\alpha_{1}(s',t')p')^{n+\nu/2}+(n+\nu/2) (\alpha_{1}(s',t')p')^{n+\nu/2-1}(\alpha_{2}(s',t')p'^2)\\&=
	(\alpha_{1}(s',t')p')^{n+\nu/2}+(n+\nu/2) (\alpha_{1}(s',t'))^{n+\nu/2-1}\alpha_{2}(s',t')p'^{n+\nu/2+1}.
\end{align}
Substituting the above approximation in the series representation of the Bessel function yields 
\begin{align}
\nonumber	&\sum\limits_{n=0}^{\infty}\frac{(-1)^nk^{2n+\nu}}{n!(n+\nu)! 2^{2n+\nu}}\left(\alpha_{1}(s',t')p'+\alpha_{2}(s',t')p'^2\right)^{n+\nu/2}\\&\simeq 	\sum\limits_{n=0}^{\infty}\frac{(-1)^nk^{2n+\nu}}{n!(n+\nu)! 2^{2n+\nu}}(	(\alpha_{1}(s',t')p')^{n+\nu/2}+(n+\nu/2) (\alpha_{1}(s',t'))^{n+\nu/2-1}\alpha_{2}(s',t')p'^{n+\nu/2+1}).
\end{align}
The first summation simply yields a single Bessel function,
\begin{align}
	\sum\limits_{n=0}^{\infty}\frac{(-1)^nk^{2n+\nu}}{n!(n+\nu)! 2^{2n+\nu}}	(\alpha_{1}(s',t')p')^{n+\nu/2}=J_\nu(k\sqrt{\alpha_{1}(s',t')p'})= J_\nu^{(l)}(k\omega).
\end{align}
The second summation gives
\begin{align}
\nonumber	&\sum\limits_{n=0}^{\infty}\frac{(-1)^nk^{2n+\nu}}{n!(n+\nu)! 2^{2n+\nu}}(	(n+\nu/2) (\alpha_{1}(s',t'))^{n+\nu/2-1}\alpha_{2}(s',t')p'^{n+\nu/2+1}\\&= \frac{\alpha_2(s',t')p'}{2\alpha_1(s',t')}\left(\nu J_\nu(k \sqrt{\alpha_1(s',t')p'})-k\sqrt{\alpha_1(s',t')p'}J_{\nu+1}(k \sqrt{\alpha_1(s',t')p'})\right)\\&=J_\nu^{(sl)}(k\omega).
\end{align}
The superscripts $(l), (sl)$ denote the approximation to leading-- and subleading--order.
Adding both, we conclude
\begin{align}
\nonumber	&J_{\nu}\left(k \left(\sum\limits_{j=1}^{\infty}\alpha_{j}(s',t')p'^j\right)^{1/2}\right)\simeq J_\nu(k\sqrt{\alpha_{1}(s',t')p'})\\&+\frac{\alpha_2(s',t')p'}{2\alpha_1(s',t')}\left(\nu J_\nu(k \sqrt{\alpha_1(s',t')p'})-k\sqrt{\alpha_1(s',t')p'}J_{\nu+1}(k \sqrt{\alpha_1(s',t')p'})\right)
\end{align}
Thus, to get the subleading--order asymptotes in inverse powers of $\Xi$, we have to effectively calculate Laplace transforms of the above expressions multiplied with the Taylor expansion around $p'=0$ of the corresponding prefactors of each Bessel function, yielding a two--dimensional integral in $s',t'$, which is evaluated exactly.
\section{Derivation of the Gaussian}

First, we calculate the second-order Taylor expansion of $\omega^2$ in \cref{eq:Rsk}, 
\begin{align}
	\omega^2\simeq \frac{2p' s'^2t'^2}{(g_a^++1)(g_b^++1)} +\frac{(g_a^+g_b^+-g_a^+-g_b^+-3)p'^2s'^4t'^4}{(g_a^++1)^2(g_b^++1)^2}.
\end{align}
As seen in the above equation, we have functional dependencies from the transmission coefficients as 
\begin{align}
	g_a^+g_b^+-g_a^+-g_b^+-3 &= \frac{4}{T_a T_b}(1-T_a-T_b),\\
	(g_a^++1)(g_b^++1) &= \frac{4}{T_aT_b}
\end{align}
Second, all the prefactors of the Bessel functions in \cref{eq:Rsk} are also written as a Taylor series up to subleading--order in $p'$.\ 
 Having all relevant expansions and hand, we now derive the universal Gaussian for $\beta=4$.
To derive the leading-order term of the asymptotic expansions, the Bessel functions of order $0$ and $2$ are replaced according to 
\begin{align}
	J_\nu(k \omega)\simeq J_\nu\left(k \sqrt{\frac{2p's'^2t'^2}{(g_a^++1)(g_b^++1)}}\right).
\end{align}
By doing so, in the leading--order approximation we have to calculate the integral 
\begin{align}
	R_s^{(l)}(k)&= 1+ \int\limits_{0}^{\infty}dp'\int\limits_{0}^{\sqrt{1/2}}ds'\int\limits_{0}^{1}dt'\exp{(-\pi\Xi p')}\Biggl(\frac{k^4 p' s'^2(2s'^2-1)t'^3}{4(g_a^++1)^2(g_b^++1)^2(s'^2-1)\sqrt{(1-t'^2)(1-s'^2(1+t'^2))}}\\&\times \Biggl(J_0\left(k \sqrt{\frac{2p's'^2t'^2}{(g_a^++1)(g_b^++1)}}\right)+J_2\left(k \sqrt{\frac{2p's'^2t'^2}{(g_a^++1)(g_b^++1)}}\right)\Biggr)\\&+ \frac{k^2 (2s'^2-1)t'^3}{2(g_a^++1)(g_b^++1)(1-s'^2)^2\sqrt{(1-t'^2)(1-s'^2(1+t'^2))}}J_0\left(k \sqrt{\frac{2p's'^2t'^2}{(g_a^++1)(g_b^++1)}}\right)\Biggr).
\end{align}
First, the $p'$--integration is done. 
The Laplace transforms are calculated by making use of the series representation of the Bessel function and yield 
\begin{align}
	\int\limits_{0}^{\infty}dp'\exp{(-\pi\Xi p')}J_0\left(k \sqrt{\frac{2p's'^2t'^2}{(g_a^++1)(g_b^++1)}}\right) &= \frac{2}{(2\pi\Xi)}\exp{\left(-\frac{k^2s'^2t'^2}{2\pi\Xi(g_a^++1)(g_b^++1)}\right)},\\
	\int\limits_{0}^{\infty}dp' \exp{(-\pi\Xi p')}p'J_2\left(k \sqrt{\frac{2p's'^2t'^2}{(g_a^++1)(g_b^++1)}}\right) &= \frac{4k^2s'^2t'^2}{(2\pi\Xi)^3(g_a^++1)(g_b^++1)}\exp{\left(-\frac{k^2s'^2t'^2}{2\pi\Xi(g_a^++1)(g_b^++1)}\right)},\\
	\int\limits_{0}^{\infty}dp'\exp{(-\pi\Xi p')} p'J_0\left(k \sqrt{\frac{2p's'^2t'^2}{(g_a^++1)(g_b^++1)}}\right) &\nonumber= \exp{\left(-\frac{k^2s'^2t'^2}{2\pi\Xi(g_a^++1)(g_b^++1)}\right)}\\&\times\Biggl(\frac{4(2\pi\Xi)(g_a^++1)(g_b^++1)-4 k^2 s'^2t'^2}{(2\pi\Xi)^3(g_a^++1)(g_b^++1)}\Biggr)
\end{align}
Setting in the result of the Laplace transforms and further algebraic simplifications yield the two-dimensional integral 
\begin{align}
	&R_s^{(l)}(k)\\&=1+ \int\limits_{0}^{\sqrt{1/2}}ds'\int\limits_{0}^{1}dt'\Biggl(\exp{\left(-\frac{k^2s'^2t'^2}{2\pi\Xi(g_a^++1)(g_b^++1)}\right)}\\&\times\frac{k^2(2s'^2-1)t'^3(k^2s'^2(s'^2-1)+(2\pi\Xi)(g_a^++1)(g_b^++1))}{(g_a^++1)^2(g_b^++1)^2(1-s'^2)^2\sqrt{(1-s'^2)(1-s'^2(1+t'^2))}(2\pi\Xi)^2}\Biggr),
\end{align}
which we evaluate exactly. To do so, it is crucial to reparameterize the $s',t'$--variables as 

\begin{align}
	s' = \sqrt{\frac{1-\sqrt{(1+x(1-y))y}}{2}},\ t' = \sqrt{1-\frac{(\sqrt{(1+x(1-y))}-\sqrt{y})^2}{2\sqrt{1-\sqrt{(1+x(1-y))y}}}}.
\end{align}
The corresponding Jacobian is 
\begin{align}
	\left|\frac{\partial(s',t')}{\partial{(x,y)}}\right| = \frac{(1+x)(1-y)^2}{16(1-\sqrt{y(1+x-x y)})\sqrt{y(1-x)(1-y)(1-x(y-1))}}.
\end{align}
This parametrization maps the integration domain to the rectangle $x\in [-1,1],\ y\in [0,1]$. 
Upon further algebraic simplification and incorporating the Jacobian of the above transformation, we arrive at 
\begin{align}
	R_s^{(l)}(k)=&\nonumber1+\int\limits_{-1}^{1}dx\int\limits_{0}^{1}dy\exp{\left(-\frac{k^2(1-x)(1-y)}{4(2\pi\Xi)(g_a^++1)(g_b^++1)}\right)}\\&\times\frac{k^2(x-1)(-k^2(1-y)(1-xy)+4(2\pi\Xi)(g_a^++1)(g_b^++1))}{16(2\pi\Xi)^2(g_a^++1)^2(g_b^++1)^2(1-xy)^2}.
\end{align}Expecting a solution to the integral with a Gaussian--like factor times a rational function in $k$ and all transmission coefficients, we make an ansatz for the $y$--antiderivative of the form \begin{align}\exp{\left(-\frac{k^2(1-x)(1-y)}{4(2\pi\Xi)(g_a^++1)(g_b^++1)}\right)}\phi(x,y),\end{align} where $\phi(x,y)$ is a rational function both in $x$ and $y$. The antiderivative, by virtue of the quotient rule, satisfies 
\begin{align}
	&\frac{\partial}{\partial y}\exp{\left(-\frac{k^2(1-x)(1-y)}{4(2\pi\Xi)(g_a^++1)(g_b^++1)}\right)}\phi(x,y) \\&= -\frac{k^2 (x-1)}{4(2\pi\Xi)(g_a^++1)(g_b^++1)}\exp{\left(-\frac{k^2(1-x)(1-y)}{4(2\pi\Xi)(g_a^++1)(g_b^++1)}\right)}\phi(x,y) \\&+\exp{\left(-\frac{k^2(1-x)(1-y)}{4(2\pi\Xi)(g_a^++1)(g_b^++1)}\right)}\frac{\partial}{\partial y}\phi(x,y) \\&=\exp{\left(-\frac{k^2(1-x)(1-y)}{4(2\pi\Xi)(g_a^++1)(g_b^++1)}\right)}\frac{k^2 (x-1)(-k^2(1-y)(1-xy)+4(2\pi\Xi)(g_a^++1)(g_b^++1))}{16(2\pi\Xi)^2(g_a^++1)^2(g_b^++1)^2(1-xy)^2}.
\end{align}
Dividing by the exponential functions on both sides, we are left with the ordinary differential equation
\begin{align}
	-\frac{k^2 (x-1)}{4(2\pi\Xi)(g_a^++1)(g_b^++1)}\phi(x,y) +\frac{\partial}{\partial y}\phi(x,y)=\frac{k^2 (x-1)(-k^2(1-y)(1-xy)+4(2\pi\Xi)(g_a^++1)(g_b^++1))}{16(2\pi\Xi)^2(g_a^++1)^2(g_b^++1)^2(1-xy)^2}.
\end{align}
By making a Möbius-type ansatz,
\begin{align}
	\phi(x,y)=\frac{A(x)y+B}{C(x)y+D},
\end{align} where the coefficients $A,\ B,\ C,\ D$ are possibly functions of $x,\ T_c$ and $k$, we arrive at 
\begin{align}
	\phi(x,y) = \frac{k^2(1-y)}{4(2\pi\Xi)(g_a^++1)(g_b^++1)(1-xy)}.
\end{align}
It should be noted that the denominator in the above expression can evaluate to zero when $x=y=1$. This singularity is integrable, since the factor $(1-y)$ of the nominator cancels the $(1-y)^{-1}$ contribution when $x=1$. Thus, we can proceed by the fundamental theorem of calculus for the iterated integrals.
Therefore, we have 
\begin{align}
	\int\limits_{0}^{1}&dy\exp{\left(-\frac{k^2(1-x)(1-y)}{4(2\pi\Xi)(g_a^++1)(g_b^++1)}\right)}\frac{k^2(x-1)(-k^2(1-y)(1-xy)+4(2\pi\Xi)(g_a^++1)(g_b^++1))}{16(2\pi\Xi)^2(g_a^++1)^2(g_b^++1)^2(1-xy)^2} \\&= \exp{\left(-\frac{k^2(1-x)(1-y)}{4(2\pi\Xi)(g_a^++1)(g_b^++1)}\right)}\frac{k^2(1-y)}{4(2\pi\Xi)(g_a^++1)(g_b^++1)(1-xy)}\Biggl|_{0}^{1} \\&= -\exp{\left(-\frac{k^2(1-x)}{4(2\pi\Xi)(g_a^++1)(g_b^++1)}\right)}\frac{k^2}{4(2\pi\Xi)(g_a^++1)(g_b^++1)}.
\end{align}The remaining integration is trivial and yields 
\begin{align}
	-\int\limits_{-1}^{1}&dx\exp{\left(-\frac{k^2(1-x)}{4(2\pi\Xi)(g_a^++1)(g_b^++1)}\right)}\frac{k^2}{4(2\pi\Xi)(g_a^++1)(g_b^++1)}= -\exp{\left(-\frac{k^2(1-x)}{4(2\pi\Xi)(g_a^++1)(g_b^++1)}\right)}\Biggl|_{-1}^{1}\\&= -1+\exp{\left(-\frac{k^2}{2(2\pi\Xi)(g_a^++1)(g_b^++1)}\right)}.
\end{align}Altogether, we have in leading--order
\begin{align}
	R_s^{(l)}(k)= 1-1+\exp{\left(-\frac{k^2}{2(2\pi\Xi)(g_a^++1)(g_b^++1)}\right)} =\exp{\left(-\frac{k^2}{2(2\pi\Xi)(g_a^++1)(g_b^++1)}\right)}.  
\end{align}
As seen from the above calculations, the Efetov-Wegner-contribution, which is given by $1$ here, gets canceled by the boundary term of the last integration. Therefore, the $\delta$-type singularity in the expression for the distributions $P_s(x_s)$ arises merely due to individually calculating the inverse Fourier transforms of the Efetov-Wegner-contribution and the integral over singular values. Compared to the  unitary case, an additional factor of $2$ arises in the denominator of the exponential. The distribution $P_s^{(l)}(x_s)$ follows by inverse Fourier transform
\begin{align}
	P_s^{(l)}(x_s) &=  \mathcal{F}^{-1}\left[ R_s(k)\right](x_s) = \sqrt{\frac{(2\pi\Xi)(g_a^++1)(g_b^++1)}{2\pi}}\exp{\left(-\frac{(2\pi\Xi)(g_a^++1)(g_b^++1)x_s^2}{2}\right)}.
\end{align}
Now, by rescaling according to $x_s= \xi_s/\sqrt{2\Xi}$, we find the universal Gaussian
	\begin{align}
	P_s^{(l)}(\xi_s) = \sqrt{\frac{ (g_a^++1)(g_b^++1)}{2}}\exp{\Biggl(-\frac{\pi (g_a^++1)(g_b^++1)\xi_s^2}{2}\Biggr)}. 
\end{align}

\section{Corrections of higher order}
To extend our analysis beyond the Gaussian approximation, we have to include higher order terms of the Taylor series of $\omega^2$ as well as the prefactors of all Bessel functions. We now have to distinguish between leading-- and subleading--order terms of prefactors and Bessel functions. All prefactors are expanded in a Taylor series around $p'=0$ up to subleading order. The subleading--order correction to the Bessel functions read
\begin{align}
	J_0^{(sl)}(k\omega) &= -\frac{(g_a^+g_b^+-g_a^+-g_b^+-3)k p'^{3/2}s'^2t'^2}{2\sqrt{2}(g_a^++1)(g_b^++1)}\sqrt{\frac{s'^2t'^2}{(g_a^++1)(g_b^++1)}}J_1\left(k \sqrt{\frac{2p's'^2t'^2}{(g_a^++1)(g_b^++1)}}\right),\\ J_2^{(sl)}(k\omega) &\nonumber=
	\frac{(g_a^+g_b^+-g_a^+-g_b^+-3)p's'^2t'^2}{4(g_a^++1)(g_b^++1)}\Biggl(2J_2\Biggl(k \sqrt{\frac{2p's'^2t'^2}{(g_a^++1)(g_b^++1)}}\Biggr)\\&- k \sqrt{\frac{2p's'^2t'^2}{(g_a^++1)(g_b^++1)}}J_3\Biggl(k \sqrt{\frac{2p's'^2t'^2}{(g_a^++1)(g_b^++1)}}\Biggr) \Biggr).
\end{align}
Further corrections are now obtained by 
\begin{align}
	R_s(k)&\nonumber\simeq \exp{\left(-\frac{k^2}{2(2\pi\Xi)(g_a^++1)(g_b^++1)}\right)}\\&\nonumber+ \int\limits_{0}^{\infty}dp'\int\limits_{0}^{\sqrt{1/2}}ds'\int\limits_{0}^{1}dt'\exp{(-\pi\Xi p')}\\&\times\Biggl( \frac{k^4 p's'^2(2s'^2-1)t'^3}{4(g_a^++1)^2(g_b^++1)^2(s'^2-1)\sqrt{(1-t'^2)(1-s'^2(1+t'^2))}}(J_0^{(sl)}(k\omega)+J_2^{(sl)}(k\omega)) \\&\nonumber+ \frac{k^4p'^2s'^2t'^3(2s'^2-1)(7-g_a^+g_b^++3(g_a^++g_b^+)+8(g_a^+g_b^+-g_a^+-g_b^+-3)s'^2t'^2)}{16(g_a^++1)^3(g_b^++1)^3(s'^2-1)\sqrt{(1-t'^2)(1-s'^2(1+t'^2))}}(J_0^{(l)}(k\omega)+J_2^{(l)}(k\omega)) \\&\nonumber+\frac{k^2 (2s'^2-1)t'^3}{2(g_a^++1)(g_b^++1)(1-s'^2)^2\sqrt{(1-t'^2)(1-s'^2(1+t'^2))}}J_0^{(sl)}(k\omega)\\&+ \frac{k^2 p'}{8}\Biggl(\frac{(2s'^2-1)t'^3}{(g_a^++1)(g_b^++1)(1-s'^2)^2\sqrt{(1-t'^2)(1-s'^2(1+t'^2))}}\Biggr)J_0^{(l)}(k\omega)\Biggr). 
\end{align}
From the above equations we infer that we need to find the following Laplace transforms of the functions $p' (J_0^{(sl)}(k\omega)+J_2^{(sl)}(k\omega)) $, $p'^2 (J_0^{(
	l)}(k\omega)+J_2^{(l)}(k\omega)) $, $J_0^{(sl)}(k\omega)$ and $p' J_0^{(l)}(k\omega)$. All of these are derived via the series representation of the Bessel function and interchanging summation and integration. After carrying out the Laplace transforms, we arrive at
\begin{align}
	\int\limits_{0}^{\infty}dp'p' \exp{(-\pi\Xi p')}(J_0^{(sl)}(k\omega)+J_2^{(sl)}(k\omega))  &\nonumber= -\frac{4(g_a^+g_b^+-g_a^+-g_b^+-3)}{(g_a^++1)^3(g_b^++1)^3(2\pi\Xi)^5} \exp{\left(-\frac{k^2 s'^2t'^2}{(2\pi\Xi)(g_a^++1)(g_b^++1)}\right)}
	\\&\times k^2 s'^4t'^4(-k^2s'^2t'^2+3(2\pi\Xi)(g_a^++1)(g_b^++1))
	, \\
	\int\limits_{0}^{\infty}dp'\exp{(-\pi\Xi p')}p'^2 (J_0^{(l)}(k\omega)+J_2^{(l)}(k\omega))  &\nonumber=\exp{\left(-\frac{k^2 s'^2t'^2}{(2\pi\Xi)(g_a^++1)(g_b^++1)}\right)}\\&\times\left(\frac{-8 k^2s'^2t'^2+16(2\pi\Xi)(g_a^++1)(g_b^++1)}{(g_a^++1)(g_b^++1)(2\pi\Xi)^4}\right),\\
	\int\limits_{0}^{\infty}dp'\exp{(-\pi\Xi p')} J_0^{(sl)}(k\omega)&\nonumber=- \exp{\left(-\frac{k^2 s'^2t'^2}{(2\pi\Xi)(g_a^++1)(g_b^++1)}\right)}\\&\times\left(\frac{2(g_a^+g_b^+-g_a^+-g_b^+-3)k^2s'^4t'^4(-k^2s'^2t'^2+2(2\pi\Xi)(g_a^++1)(g_b^++1))}{(g_a^++1)^3(g_b^++1)^3(2\pi\Xi)^4}\right)\\
	\int\limits_{0}^{\infty}dp'\exp{(-\pi\Xi p')} p' J_0^{(l)}(k\omega)  &=\exp{\left(-\frac{k^2 s'^2t'^2}{(2\pi\Xi)(g_a^++1)(g_b^++1)}\right)}\\&\times\frac{4(2\pi\Xi)(g_a^++1)(g_b^++1)-4k^2s'^2t'^2}{(g_a^++1)(g_b^++1)(2\pi\Xi)^3}.
\end{align}
Setting these results into the integral from above, the resulting two-dimensional integral for the subleading--order correction reads 
\begin{align}
	R_s^{(sl)}(k)&\nonumber=\int\limits_{0}^{\sqrt{1/2}}ds'\int\limits_{0}^{1}dt'\Biggl( \frac{k^4 s'^2(2s'^2-1)t'^3}{4(g_a^++1)^2(g_b^++1)^2(s'^2-1)\sqrt{(1-t'^2)(1-s'^2(1+t'^2))}}\Biggl(-\frac{4(g_a^+g_b^+-g_a^+-g_b^+-3)}{(g_a^++1)^3(g_b^++1)^3(2\pi\Xi)^5}\\&\nonumber\times \exp{\left(-\frac{k^2 s'^2t'^2}{(2\pi\Xi)(g_a^++1)(g_b^++1)}\right)}k^2 s'^4t'^4(-k^2s'^2t'^2+3(2\pi\Xi)(g_a^++1)(g_b^++1))\Biggr) \\&\nonumber+ \frac{k^4s'^2t'^3(2s'^2-1)(7-g_a^+g_b^++3(g_a^++g_b^+)+8(g_a^+g_b^+-g_a^+-g_b^+-3)s'^2t'^2)}{16(g_a^++1)^3(g_b^++1)^3(s'^2-1)\sqrt{(1-t'^2)(1-s'^2(1+t'^2))}}\\&\nonumber\times\Biggl(\exp{\left(-\frac{k^2 s'^2t'^2}{(2\pi\Xi)(g_a^++1)(g_b^++1)}\right)}\left(\frac{-8 k^2s'^2t'^2+16(2\pi\Xi)(g_a^++1)(g_b^++1)}{(g_a^++1)(g_b^++1)(2\pi\Xi)^4}\right)\Biggr) \\&\nonumber+\frac{k^2 (2s'^2-1)t'^3}{2(g_a^++1)(g_b^++1)(1-s'^2)^2\sqrt{(1-t'^2)(1-s'^2(1+t'^2))}}\\&\nonumber\times\Biggl(- \exp{\left(-\frac{k^2 s'^2t'^2}{(2\pi\Xi)(g_a^++1)(g_b^++1)}\right)}\\&\nonumber\times\left(\frac{2(g_a^+g_b^+-g_a^+-g_b^+-3)k^2s'^4t'^4(-k^2s'^2t'^2+2(2\pi\Xi)(g_a^++1)(g_b^++1))}{(g_a^++1)^3(g_b^++1)^3(2\pi\Xi)^4}\right)\Biggr)\\&\nonumber+ \frac{k^2 }{8}\Biggl(\frac{(2s'^2-1)t'^3}{(g_a^++1)(g_b^++1)(1-s'^2)^2\sqrt{(1-t'^2)(1-s'^2(1+t'^2))}}\Biggr)\\&\times\Biggl(\exp{\left(-\frac{k^2 s'^2t'^2}{(2\pi\Xi)(g_a^++1)(g_b^++1)}\right)}\frac{4(2\pi\Xi)(g_a^++1)(g_b^++1)-4k^2s'^2t'^2}{(g_a^++1)(g_b^++1)(2\pi\Xi)^3}\Biggr)\Biggr).
\end{align}
Applying the same transformation as in the derivation of the Gaussian, the above integral is equivalent to 
\begin{align}
R_s^{(sl)}(k)=	&\nonumber\int\limits_{-1}^{1}\int\limits_{0}^{1}dxdy
	\Biggl(\exp{\left(-\frac{k^2(1-x)(1-y)}{4(2\pi\Xi)(g_a^++1)(g_b^++1)}\right)}\\\nonumber&\times\frac{ k^2 (x -1) (-k^2 (y-1) (x y-1) + 4 (1 + g_a^+) (1 + g_b^+)(2\pi\Xi))}{1024 (1 + g_a^+)^5 (1 + g_b^+)^5 (x y-1)^2 (2\pi\Xi)^5}\\&\nonumber\times ((g_a^+g_b^+-g_a^+-g_b^+-3) k^4 (x -1)^3 (y-1)^3 \\&\nonumber- 
	8 (1 + g_a^+) (1 + g_b^+) k^2 (x -1) (y-1) ((1 + g_a^+) (1 + g_b^+) + 
	2 (3 + g_a^+ + g_b^+ - g_a^+ g_b^+) x \\&\nonumber+ 
	2 (g_a^+g_b^+-g_a^+-g_b^+-3) (x -1) y) (2\pi\Xi) \\&\nonumber+ 
	32 (1 + g_a^+)^2 (1 + 
	g_b^+)^2 (2 (2 + g_a^+ + g_b^+) \\&\nonumber+ (3 + g_a^+ + g_b^+ - g_a^+ g_b^+) x + (3 + g_a^+ + g_b^+ - 
	g_a^+ g_b^+) y + (g_a^+g_b^+-g_a^+-g_b^+-3) x y) (2\pi\Xi)^2) \\&\nonumber+ 	\exp{\left(-\frac{k^2(1-x)(1-y)}{4(2\pi\Xi)(g_a^++1)(g_b^++1)}\right)}\frac{k^2(y-1)}{256(g_a^++1)^4(g_b^++1)^4(x y-1)(2\pi\Xi)^4}\\&\nonumber\times(3 (g_a^+g_b^+-g_a^+-g_b^+-3) k^4 (x -1)^3 (y-1)^2 \\&\nonumber- 
	16 (1 + g_a^+) (g_a^+g_b^+-g_a^+-g_b^+-3) (1 + g_b^+) k^2 (x -1)^2 (-1 + 
	y)(2\pi\Xi) \\&\nonumber- 
	8 (1 + g_a^+) (1 + g_b^+) k^2 (x -1) ((1 + g_a^+) (1 + g_b^+) + 
	2 (3 + g_a^+ + g_b^+ - g_a^+ g_b^+) x \\&\nonumber+ 
	2 (g_a^+g_b^+-g_a^+-g_b^+-3) (x -1) y)(2\pi\Xi) \\&+ 
	32 (1 + g_a^+)^2 (1 + g_b^+)^2 (3 + g_a^+ + g_b^+ - 
	g_a^+ g_b^+ + (g_a^+g_b^+-g_a^+-g_b^+-3) x)(2\pi\Xi)^2)\Biggr).
\end{align}
Again, we solve the $y$-integral via a first-order differential equation and arrive at 
\begin{align}
	R_s^{(sl)} (k)&\nonumber= \int\limits_{-1}^{1}dx \exp{\left(-\frac{k^2(1-x)(1-y)}{4(2\pi\Xi)(g_a^++1)(g_b^++1)}\right)}\frac{k^2(y-1)}{256(g_a^++1)^4(g_b^++1)^4(x y-1)(2\pi\Xi)^4}\\&\nonumber\times(((g_a^+g_b^+-g_a^+-g_b^+-3) k^4 (x-1)^3 (y -1)^3 \\&\nonumber- 
	8 (1 + g_a^+) (1 + g_b^+) k^2 (x-1) (y -1) ((1 + g_a^+) (1 + g_b^+) \\&\nonumber+ 
	2 (3 + g_a^+ + g_b^+ - g_a^+ g_b^+) x + 
	2 (g_a^+g_b^+-g_a^+-g_b^+-3) (x-1) y)(2\pi\Xi) \\&\nonumber+ 
	32 (1 + g_a^+)^2 (1 + 
	g_b^+)^2 (2 (2 + g_a^+ + g_b^+) + (3 + g_a^+ + g_b^+ - g_a^+ g_b^+) x \\&+ (3 + g_a^+ + g_b^+ - 
	g_a^+ g_b^+) y + (g_a^+g_b^+-g_a^+-g_b^+-3) x y)(2\pi\Xi)^2))\Biggl|_{0}^{1}\\&\nonumber= -\int\limits_{-1}^{1}dx \frac{k^2}{256(g_a^++1)^4(g_b^++1)^4(2\pi\Xi)^4}\exp{\left(-\frac{k^2(1-x)}{4(2\pi\Xi)(g_a^++1)(g_b^++1)}\right)}\\&\nonumber\times 
	(-((g_a^+g_b^+-g_a^+-g_b^+-3) k^4 (x-1)^3) + 
	8 (1 + g_a^+) (1 + g_b^+) k^2 (x-1) ((1 + g_a^+) (1 + g_b^+) \\&\nonumber+ 
	2 (3 + g_a^+ + g_b^+ - g_a^+ g_b^+) x) (2\pi\Xi) + 
	32 (1 + g_a^+)^2 (1 + 
	g_b^+)^2 (2 (2 + g_a^+ + g_b^+) \\&+ (3 + g_a^+ + g_b^+ - g_a^+ g_b^+) x) (2\pi\Xi)^2).
\end{align}
The remaining integral is done by direct integration and yields 
\begin{align}
	R_s^{(sl)} (k)=	&\nonumber-\int\limits_{-1}^{1}dx \frac{k^2}{256(g_a^++1)^4(g_b^++1)^4(2\pi\Xi)^4}\exp{\left(-\frac{k^2(1-x)}{4(2\pi\Xi)(g_a^++1)(g_b^++1)}\right)}\\&\nonumber\times 
	(-((g_a^+g_b^+-g_a^+-g_b^+-3) k^4 (-1 + x)^3) + 
	8 (1 + g_a^+) (1 + g_b^+) k^2 (-1 + x) ((1 + g_a^+) (1 + g_b^+) \\&\nonumber+ 
	2 (3 + g_a^+ + g_b^+ - g_a^+ g_b^+) x) (2\pi\Xi) + 
	32 (1 + g_a^+)^2 (1 + 
	g_b^+)^2 (2 (2 + g_a^+ + g_b^+) \\&\nonumber+ (3 + g_a^+ + g_b^+ - g_a^+ g_b^+) x) (2\pi\Xi)^2)\\\nonumber&=\exp{\left(-\frac{k^2(1-x)}{4(2\pi\Xi)(g_a^++1)(g_b^++1)}\right)}k^2(x-1)\\&\nonumber\times ((g_a^+g_b^+-g_a^+-g_b^+-3) k^2 (-1 + x)^2 + 
	4 (1 + g_a^+) (1 + g_b^+) (g_a^+g_b^+-g_a^+-g_b^+-3 \\&\nonumber- 
	4 (2 + g_a^+ + g_b^+) + (-3 - g_a^+ - g_b^+ + g_a^+ g_b^+) x) (2\pi\Xi)) \frac{1}{64(g_a^++1)^3(g_b^++1)^3(2\pi\Xi)^3}\Biggl|_{-1}^{1}\\&\nonumber=\exp{\left(-\frac{k^2}{2(2\pi\Xi)(g_a^++1)(g_b^++1))}\right)}\\&\times\Biggl(-k^2\frac{2+g_a^++g_b^+}{2(g_a^++1)^2(g_b^++1)^2(2\pi\Xi)^2}+k^4\frac{g_a^+g_b^+-g_a^+-g_b^+-3}{8(g_a^++1)^3(g_b^++1)^3(2\pi\Xi)^3}\Biggr)
\end{align}
Combining the above result with the leading--order asymptote results in the approximation
\begin{align}
	R_s(k)&\simeq\nonumber\exp{\left(-\frac{k^2}{2(g_a^++1)(g_b^++1)(2\pi\Xi)}\right)}\\&\label{eq:CharakteristischeGSE}\times\Biggl(1-k^2\frac{2+g_a^++g_b^+}{2(g_a^++1)^2(g_b^++1)^2(2\pi\Xi)^2}+k^4\frac{g_a^+g_b^+-g_b^+-g_b^+-3}{8(g_a^++1)^3(g_b^++1)^3(2\pi\Xi)^3}\Biggr).
\end{align}
\cref{eq:CharakteristischeGSE} constitutes one of the key results of this paper, as all the information about the distribution of real and imaginary part of $S_{ab}$ as well as the cross--section distribution are calculated via integral transforms of the above expression. Here, we see that in the polynomial factor, in contrast to the unitary case, an additional term $\propto (-k^2)$ arises. This term narrows the characteristic function, since 
\begin{align}
\frac{2+g_a^++g_b^+}{2(g_a^++1)^2(g_b^++1)^2(2\pi\Xi)^2} >0
\end{align} for all finite values of $T_c$. This yields an additional broadening of the probability distribution $P_s(\xi_s)$ compared to the unitary case. After doing an inverse Fourier transform and rescaling we find 
\begin{align}
P_s^{(l)}(\xi_s) &=  \sqrt{\frac{ (g_a^++1)(g_b^++1)}{2}}\exp{\Biggl(-\frac{\pi (g_a^++1)(g_b^++1)\xi_s^2}{2}\Biggr)},\\
P_s^{(sl)}(\xi_s) &\nonumber=\sqrt{\frac{ (g_a^++1)(g_b^++1)}{2}}\exp{\Biggl(-\frac{\pi (g_a^++1)(g_b^++1)\xi_s^2}{2}\Biggr)}\\&\nonumber\times\Biggl(
\frac{5(g_a^+ g_b^+-g_b^+-g_b^+-3)-2(g_a^++1)(g_b^++1)}{16(g_a^++1)(g_b^++1)\pi\Xi}\\&\nonumber+\frac{2(2+g_a^++g_b^+)-3(g_a^+g_b^+-g_a^+-g_b^+-3)}{8\Xi}\xi_s^2\\&+\frac{(g_a^++1)(g_b^++1)(g_a^+g_b^+-g_a^+-g_b^+-3)}{16\Xi}\pi \xi_s^4 \Biggr).
\end{align}
In the onset of the Ericson regime, we therefore have a shift to the pure Gaussian that depends on the transmission coefficients $T_a,\ T_b$ and $\Xi$. At $\xi_s=0$ , we find
\begin{align}
P_s(0)=\sqrt{\frac{ (g_a^++1)(g_b^++1)}{2}}\left(1+\frac{5(g_a^+ g_b^+-g_b^+-g_b^+-3)-2(g_a^++1)(g_b^++1)}{16(g_a^++1)(g_b^++1)\pi\Xi}\right)+\mathcal{O}(\Xi^{-2}).
\end{align}

Finally, the moments up to subleading order in $1/\Xi$ read 
\begin{align}
	\overline{x_s^{2n}} &\nonumber= \frac{(2n)!}{n!}\Biggl(\frac{1}{(2(g_a^++1)(g_b^++1)(2\pi\Xi))^n}\\&\nonumber+
	n \frac{1}{(2(g_a^++1)(g_b^++1)(2\pi\Xi))^{n-1}}\frac{2+g_a^++g_b^+}{(2(g_a^++1)^2(g_b^++1)^2(2\pi\Xi)^2)}\\&+n(n-1)\frac{1}{(2(g_a^++1)(g_b^++1)(2\pi\Xi))^{n-2}}\frac{g_a^+g_b^+-g_a^+-g_b^+-3}{(8(g_a^++1)^3(g_b^++1)^3(2\pi\Xi)^3)}\Biggr).
\end{align}

\section{Cross-section distribution and modulus}
In the case of radially symmetric bivariate characteristic functions, as present in the $\beta=2$ and $\beta=4$  cases, the distribution of cross-sections is obtained from the bivariate characteristic function via a Hankel transform, $e.g.$
\begin{align}
	p(\sigma_{ab})&=\frac{1}{2}\mathcal{H}_0[R(K)](\sqrt{\sigma_{ab}})=\frac{1}{2}\int\limits_{0}^{\infty}dK K R(K)J_0(K\sqrt{\sigma_{ab}}).
\end{align}
As in the unitary case, the bivariate characteristic has structurally the same asymptotic form as the univariate characteristic function \cite{LetterUniversality}. We have 
\begin{align}
	p(\sigma_{ab})&\nonumber\simeq  \frac{1}{2}\int\limits_{0}^{\infty}dK K \exp{\left(-\frac{K^2}{2(g_a^++1)(g_b^++1)(2\pi\Xi)}\right)}\\&\times\Biggl(1-K^2\frac{2+g_a^++g_b^+}{2(g_a^++1)^2(g_b^++1)^2(2\pi\Xi)^2}+K^4\frac{g_a^+g_b^+-g_b^+-g_b^+-3}{8(g_a^++1)^3(g_b^++1)^3(2\pi\Xi)^3}\Biggr)J_0(K\sqrt{\sigma_{ab}}).
\end{align}
After carrying out the Hankel transform and rescaling according to $\sigma = \widetilde{\sigma}/(\pi\Xi(g_a^++1)(g_b^++1))$, the distribution of rescaled cross--section reads
\begin{align}
	p(\widetilde{\sigma}_{ab})\nonumber\simeq \exp{(-\widetilde{\sigma}_{ab})}\Biggl(1+\frac{g_a^+g_b^+-g_b^+-g_b^+-3 - (2+g_a^++g_b^+) }{2(g_a^++1)(g_b^++1)\pi\Xi}\\-\frac{2(g_a^+g_b^+-g_a^+-g_b^+-3 )- (2+g_a^++g_b^+)}{2(g_a^++1)(g_b^++1)\pi\Xi}\widetilde{\sigma}_{ab}+\frac{g_a^+g_b^+-g_a^+-g_b^+-3}{4(g_a^++1)(g_b^++1)\pi\Xi}\widetilde{\sigma}_{ab}^2\Biggr).
\end{align}

Thus, to subleading--order in the Ericson regime, we have a shift compared to the pure exponential decay such that 
\begin{align}\label{eq:shiftGSE}
	p(0)= 1+\frac{g_a^+g_b^+-g_b^+-g_b^+-3 - (2+g_a^++g_b^+) }{2(g_a^++1)(g_b^++1)\pi\Xi}+\mathcal{O}(\Xi^{-2}). 	
\end{align}
Consequently, the distribution of rescaled modulus, $\widetilde{r} = \sqrt{\widetilde{\sigma}_{ab}}$ reads 
\begin{align}\nonumber
	p(\widetilde{r})\simeq 2\widetilde{r}\exp{(-\widetilde{r}^2)}\Biggl(1+\frac{g_a^+g_b^+-g_b^+-g_b^+-3 - (2+g_a^++g_b^+) }{2(g_a^++1)(g_b^++1)\pi\Xi}\\-\frac{2(g_a^+g_b^+-g_a^+-g_b^+-3 )- (2+g_a^++g_b^+)}{2(g_a^++1)(g_b^++1)\pi\Xi}\widetilde{r}^2+\frac{g_a^+g_b^+-g_a^+-g_b^+-3}{4(g_a^++1)(g_b^++1)\pi\Xi}\widetilde{r}^4\Biggr).
\end{align}
\section{Comparison with numerical simulations}
\begin{figure}[hbpt]
	\centering
	\captionsetup[subfigure]{labelformat=empty}
	\begin{subfigure}[t]{0.45\linewidth}
		\centering
		\includegraphics[width=1\linewidth]{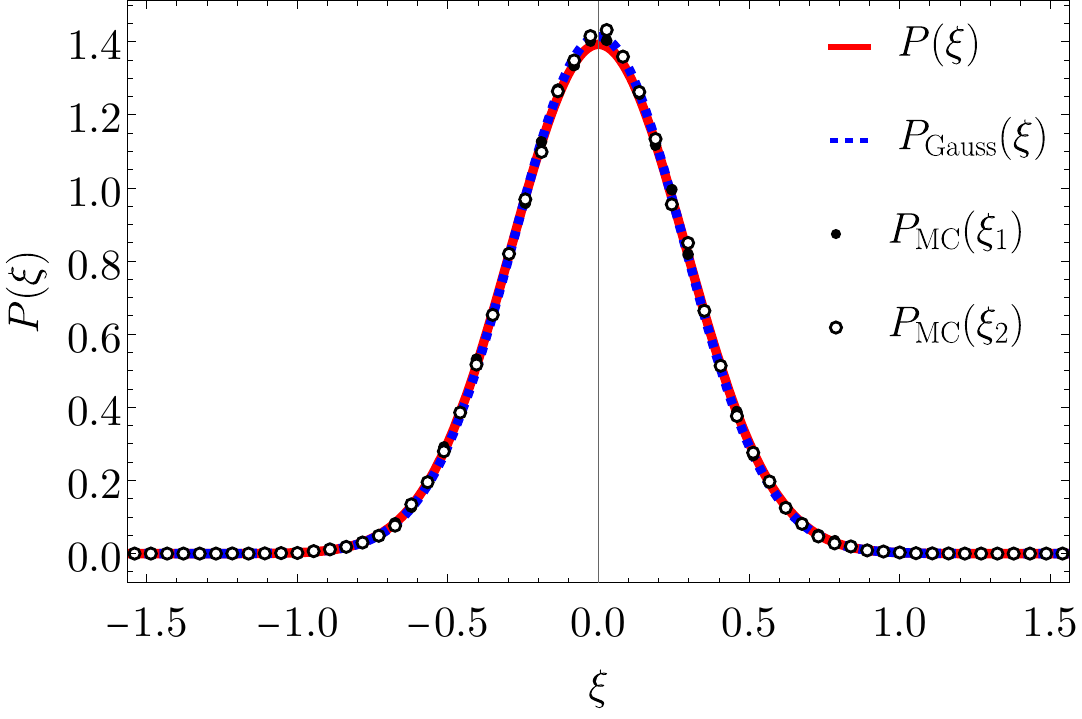} 
		\subcaption{}
		
	\end{subfigure}	
	~
	\begin{subfigure}[t]{0.45\linewidth}
		\centering
		\includegraphics[width=1\linewidth]{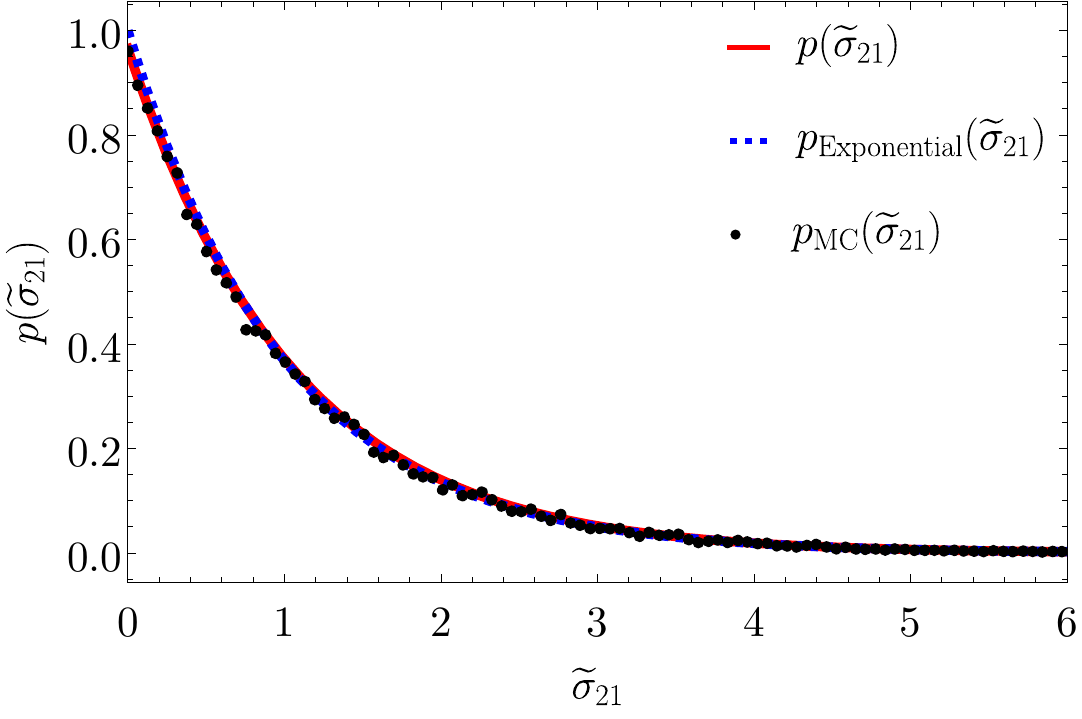} 
		\subcaption{}
		
	\end{subfigure}	
	~	\begin{subfigure}[t]{0.45\linewidth}
		\centering
		\includegraphics[width=1\linewidth]{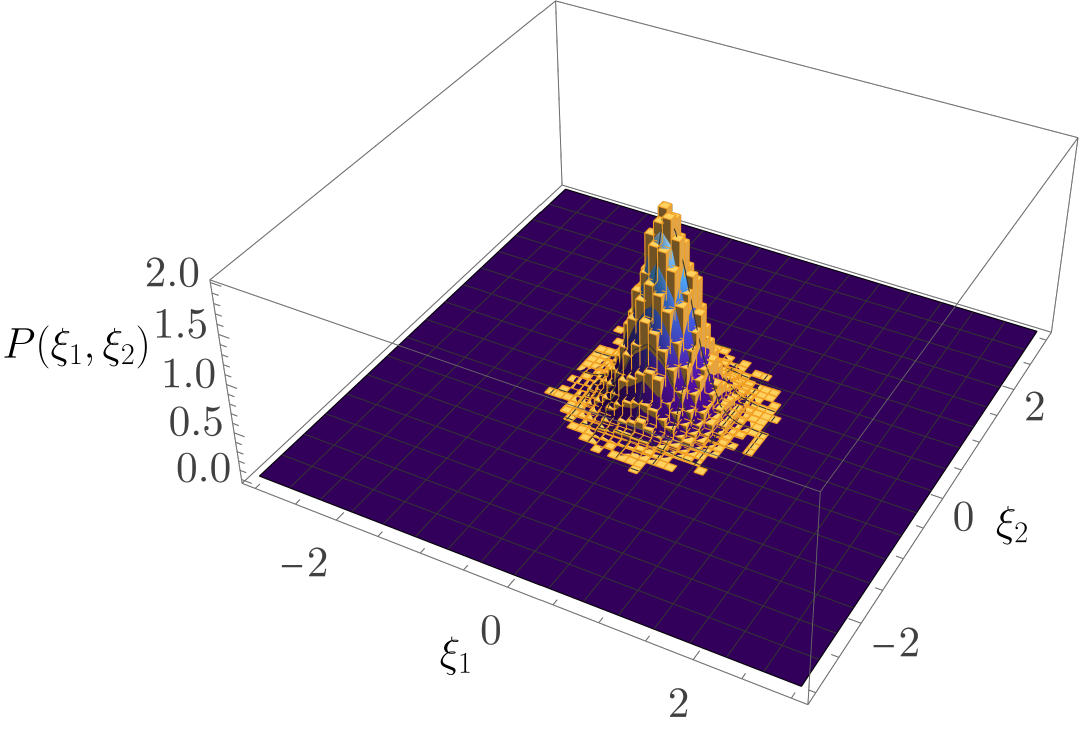} 
		\subcaption{}
		
	\end{subfigure}	
	
	\caption{Numerical simulation and the distributions of rescaled scattering matrix
		elements, cross sections and joint distribution for
		$\Xi=9.23$ and $M=60$, $T_c=1$ for all channels.}
	\label{fig:Ericson}
\end{figure}

To further validate our analytical results, we compare with Monte Carlo simulations. We draw $100000$ matrices with $N=300$ from the GSE. To look into the onset of the Ericson regime we choose $T_1=T_2=1, \ \Xi=1.57, M=30$. Thus, we have two strong channels and $28$ weak channels with $T_c=0.303\  \forall c\neq1, 2$.
\begin{figure}[htbp]
	\centering
	\captionsetup[subfigure]{labelformat=empty}
	\begin{subfigure}[t]{0.45\linewidth}
		\centering
		\includegraphics[width=\linewidth]{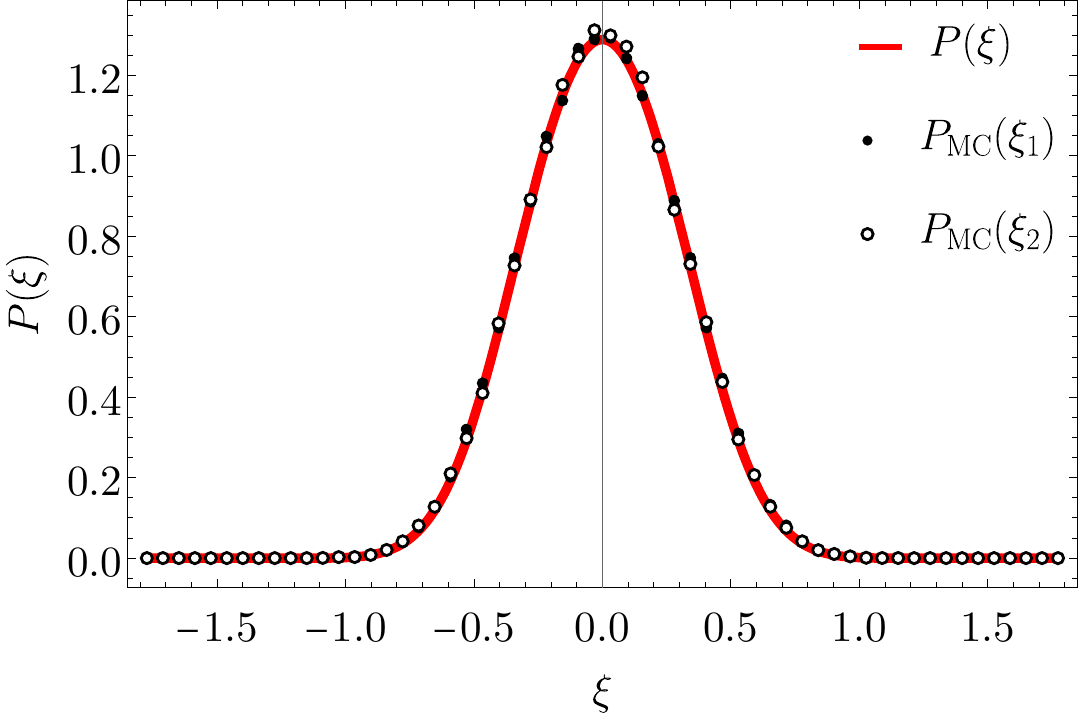} 
		\subcaption{}
	\end{subfigure}
	~
	\begin{subfigure}[t]{0.45\linewidth}
		\centering
		\includegraphics[width=1\linewidth]{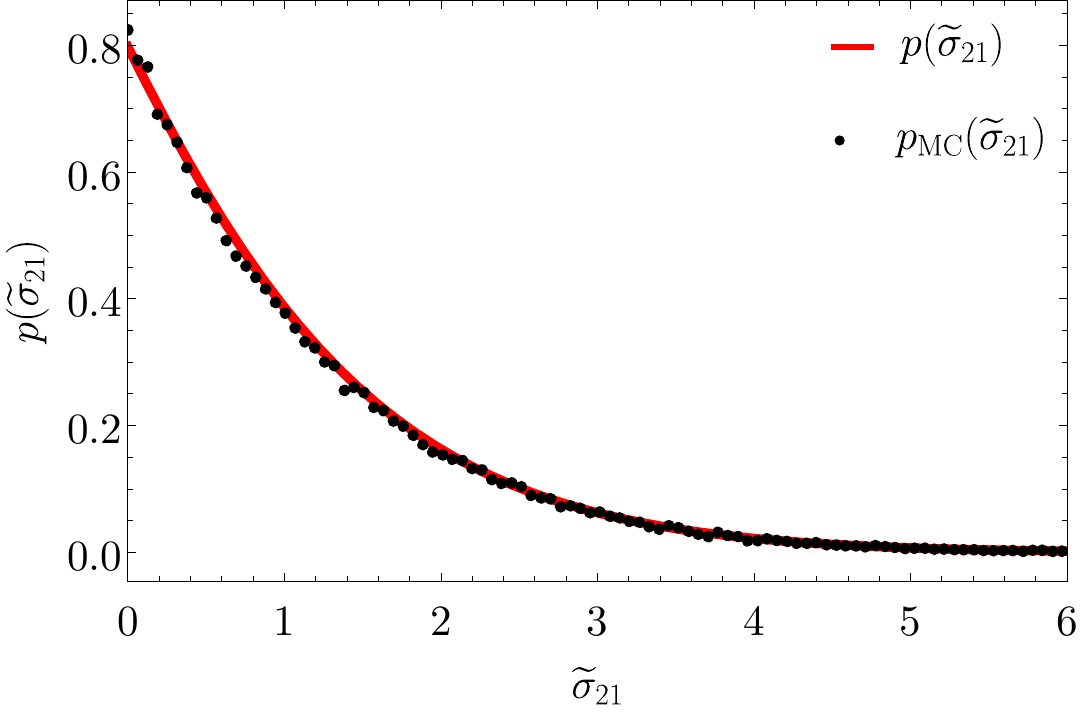} 
		\subcaption{}
		
	\end{subfigure}	
	~
	\begin{subfigure}[t]{0.45\linewidth}
		\centering
		\includegraphics[width=1\linewidth]{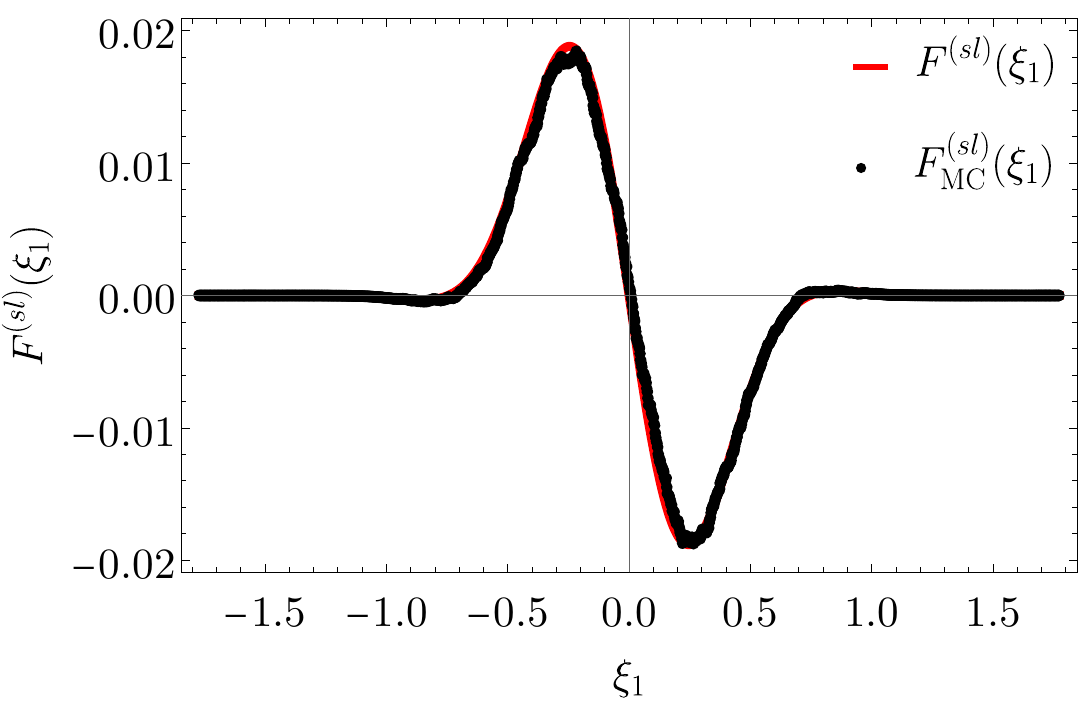} 
		\subcaption{}
		
	\end{subfigure}	
	~	\begin{subfigure}[t]{0.45\linewidth}
		\centering
		\includegraphics[width=1\linewidth]{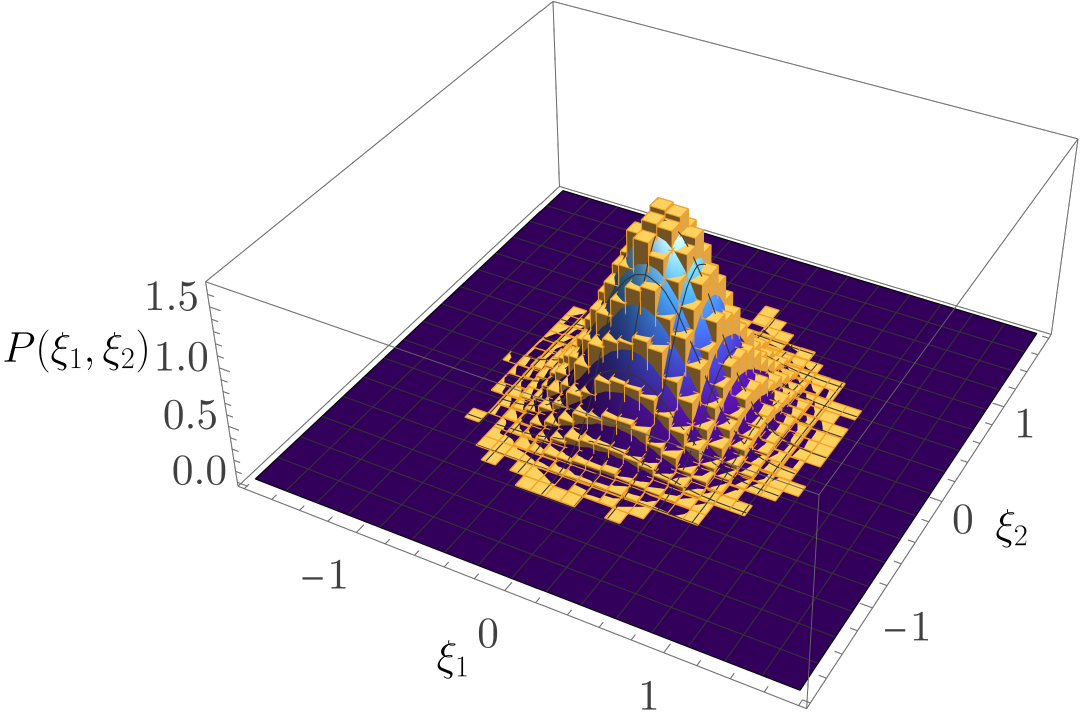} 
		\subcaption{}
		
	\end{subfigure}	
	
	\caption{Numerical simulation and the distributions of rescaled scattering matrix
		elements and cross sections (top) for
		$\Xi=1.57$ and $M=30$, $T_1=T_2=1$. Numerical simulation and subleading order term $F^{(sl)}(\xi_1)$ of the CDF and full joint distribution (bottom).}
	\label{fig:Corrections}
\end{figure}
As seen in \cref{fig:Corrections}, the analytical results are in full accordance with the Monte Carlo simulation. We observe that the distributions of rescaled real and imaginary part of $S_{ab}$ are indeed the same, which is also visualized by the radial symmetry of the joint distribution $P(\xi_1, \xi_2)$. The maximum of the rescaled distribution of cross--sections is shifted by the contribution given in \cref{eq:shiftGSE}. The subleading--order term for the CDF shows that the asymptotic correction fully captures the transition behavior even for a relatively small value of $\Xi=1.57$. This indicates that the Ericson transition is occurring really fast in the sense that higher order terms of the asymptotic expansion are almost vanishingly small when considering $\Xi>1$. This was also observed in recent experiments for the unitary case \cite{LetterUniversality}. To finalize our numerical analysis, we show that the distributions $P_s(\xi_s)$ and $p(\widetilde{\sigma}_{ab})$ turn into a Gaussian and exponential distribution, respectively, when considering $\Xi\gg 1$. We choose $M=60$ channels with perfect coupling $T_c=1$, yielding $\Xi=9.23$. We again find full agreement of the asymptotic result with the numerical simulation in \cref{fig:Ericson}.
\pagebreak
\section{Conclusion}
We were able to solve the Ericson transition for the case of time--reversal invariant systems with odd spin by means of asymptotic expansions in $1/\Xi$. As in the unitary case, we find universal Gaussian and exponential behavior of the distributions of rescaled real-- and imaginary parts of $S$--matrix elements and cross--sections. In the limit of $\Xi \gg 1$, these distributions show the same dependencies from $\Xi,\ T_a,\ T_b$ as the distributions in the unitary case. The asymptotic correction terms in the symplectic case reveal an additional structure compared to the unitary case, leading to an additional broadening of the probability distribution $P_s(\xi_s)$ and contributing a corresponding shift to the maximum of $P_s(\xi_s)$ and $p(\widetilde{\sigma}_{ab})$. We emphasize that the Ericson transition is occurring extremely fast. Numerical simulations reveal that the subleading--order correction to the CDF captures the full problem entirely even when $\Xi$ attains the small value of $\Xi=1.57$. 
\section*{Acknowledgments}
We thank Nils Gluth and Barbara Dietz for fruitful discussions. The authors are grateful to the Deutsche Forschungsgemeinschaft (DFG, German Research Foundation) for support within the project Stochastic Quantum Scattering—New Tools, New Aspects, DFG Project No. 540160740. 
\newpage
     \bibliography{EricsonTimeReversal}
\end{document}